# Multi-Class classification of vulnerabilities in Smart Contracts using AWD-LSTM, with pre-trained encoder inspired from natural language processing

Ajay K. Gogineni[1,2], S. Swayamjyoti[1,3], Devadatta Sahoo[1], Kisor K. Sahu[2,3,4]*, Raj kishore[2,3]

1. NetTantra Technologies (India) Pvt. Ltd., Bhubaneswar, Odisha 751021, India
2. Virtual and Augmented Reality Centre of Excellence, IIT Bhubaneswar 752050
3. School of Minerals, Metallurgical and Materials Engineering, IIT Bhubaneswar 752050
4. Centre of Excellence for Novel Energy Materials, IIT Bhubaneswar 752050

*Corresponding author kisorsahu@iitbbs.ac.in

Abstract

Vulnerability detection and safety of smart contracts are of paramount importance because of their immutable nature. Symbolic tools like OYENTE and MAIAN are typically used for vulnerability prediction in smart contracts. As these tools are computationally expensive, they are typically used to detect vulnerabilities until some predefined invocation depth. These tools require more search time as the invocation depth increases. Since the number of smart contracts is increasing exponentially, it is difficult to analyze the contracts using these traditional tools. Recently a machine learning technique called Long Short Term Memory (LSTM) has been used for binary classification, i.e., to predict whether a smart contract is vulnerable or not. This technique requires nearly constant search time as the invocation depth increases. In the present article, we have shown a multi-class classification, where we classify a smart contract in Suicidal, Prodigal, Greedy, or Normal categories. We used Average Stochastic Gradient Descent Weight-Dropped LSTM (AWD-LSTM), which is a variant of LSTM, to perform classification. We reduced the class imbalance (a large number of normal contracts as compared to other categories) by considering only the distinct opcode combination for normal contracts. We have achieved a weighted average $F_{beta}$ score of 90.0%. Hence, such techniques can be used to analyze a large number of smart contracts and help to improve the security of these contracts.

## 1. Introduction

"Smart contract" (SC), a term coined by Nick Szabo in 1996[1], is an extended idea of a bitcoin blockchain and consists of a bunch of promises which should be executed during transactions between mutually distrusted nodes without the mediation of any centralized trusted authority. Therefore, a smart contract is an auto executable digital technological solution, which has the potential to augment, expedite, and enhance efficacies of traditional legal contracts. While the sanctity of a traditional legal contract is ensured by institutions of law and enforcement agencies, none of them is a critical requirement for SCs, therefore significantly reducing the complexity by technological interventions. It eliminates the need for third-party entities to handle how contractual agreements (or disagreements) need to be played out between two or more parties. In real life, a vending machine is the best example of a smart contract. It eliminates the need for a shopkeeper who counts the money given by the buyer and gives the desired product. So here, only two parties are involved, the buyer and the vending machine. In SCs, the transaction between two parties is encoded in a computer code, typically using Contract Oriented Language (COL), which will be discussed in *section 1.1*. Recently, many platforms have also been used for writing smart contracts like Ethereum and Rootstock (RSK) [2-3]. The contracts for Ethereum are written in the Ethereum Virtual Machine (EVM) and programmed through a language called Solidity [4]. As these smart contracts hold virtual coins worth thousands of USD each, their safety is of paramount importance. There are some security challenges in smart contracts. First, they cannot be upgraded or patched like other traditional consumer device software after their deployment. Second, contracts are relatively difficult to test because of repeated invocations of transactions. Third, since the blockchain is dealing with cryptocurrencies worth billions of USD, they are more likely to be exploited by implanting bugs in contracts by attackers. There are few incidences of such exploitations of smart contracts. For example, in June 2016, a decentralized investment fund named DAO (Decentralised Autonomous Organisation) lost approximately $70 million due to the stealing of over 3.6 million Ether (virtual coin used by Ethereum) [5]. In November 2017, a security alert was issued by Parity Technologies, saying that their parity wallet (multi-sig wallets) was affected due to which $300 million was frozen [6]. To understand and predict such important vulnerabilities, we are resorting to the field of machine learning. Due to the ever increasing availability of high computational capabilities, machine learning methods are becoming a popular choice in analyzing data from different fields, such as analysing bio-medical image scans[7-8], Satellite images[9], materials characterization [10-11], share market[12], etc. There are several examples of the adoption of machine learning tools for different types of security attack detection. Dou *et al.* [13] used a deep LSTM model for anomaly detection in systems. Similarly, Shen *et al.* [14] used Recurrent Neural Network (RNN) having a sequence memory architecture for forecasting security events on a computer. Shin *et al.* [15] used LSTM for the identification of functions in binaries. These examples indicate that machine learning tools can be used to understand and improve the functioning of these cryptocurrency-based transactions. Let us discuss briefly some important ideas which will be useful in understanding the machine learning and smart contract fields together.

**1.1 Contract Oriented Language (COL)**

Contract oriented languages are used to write smart contracts involving various cryptocurrencies. The first COL, which is known as "Solidity," was developed by Gavin Wood, Christian Reitwiessner, Yoichi Hirai, and several of Ethereum's core contributors for writing smart contracts that are functioned on Ether [16]. Presently Solidity has worldwide more than 200,000 developers [17]. Similarly, Golang is another very popular COL. It has worldwide more than 800,000 developers, primarily because it is an open-source programming language loosely based on the syntax of the C programming language. The other COLs used for writing SCs includes JavaScript, C++, Java, SQL, FLETA, SQP, etc.

**1.2 Opcode**

Let us try to illustrate the concept of opcode by an analogy. In object-oriented programming (OOP), when we execute a computer program, it is converted into a binary representation. However, in COL, it is converted to bytecodes, which is nothing but the hexadecimal representation of the COL. The set of human-readable instructions that need to be executed in the SCs are known as opcodes. For example, in Ethereum virtual machine, the maximum number of possible distinct opcodes is only 150 [16]. All the opcodes have their hexadecimal counterparts, for example, "AND" is "0x16", "SDIV" is "0x05", "ADD" is "0x01" …etc.

There are several vulnerability detection methods in a smart contract [18,19]. Symbolic analysis, also known as automatic bug detection protocol, is currently the most widely used method for vulnerability detection. OYENTE [20, 21] and MAIAN [22,23] are the two most prominent tools employing these ideas. MAIAN is used to extract opcodes from the bytecodes. MAIAN also classifies the bytecodes and hence generates the labels required for training a neural network. The major drawback of these two tools is that, while testing a smart contract, it must be active. The other major drawback of MAIAN is that the accuracy is critically dependent on the "invocation depth," which will be discussed in the next section.

**1.3 Invocation depth**

When a user initiates a smart contract written in Solidity, the contract will be assigned a unique address and a state value. State value consists of many different fields such as Ether value, Gas points, private storage, and executable codes. Whenever we want to execute the SC, this address is used to point to this particular contract. We can do either of the two following things with a smart contract:

running the content, or, retrieving the state information from it. It is important to note at this point is that, while the execution of the smart contract, the state value is modified, but during retrieval, this remains unchanged. "Invocation depth" refers to the number of times the SC is called for [23]. The accuracy of MAIAN is dependent on the invocation depth. It means that, if we have a large invocation depth, the vulnerability present at a later invocation might not be detectable using MAIAN. To further illustrate the point, let us consider a hypothetical scenario as following. Assume a particular smart contract has more than a hundred invocation depth. Since identifying the vulnerabilities through MAIAN is resource-intensive, we might consider terminating the effort after checking up to few tens (say, 50) invocation depth. However, it might be possible that the vulnerability exists at a later instance of invocation, which will remain undetected. Some works focused on detecting vulnerabilities in smart contracts across multiple invocations [24]. The significant advantage of using an opcode based technique is that the smart contract need not be active while testing. Another major benefit of this approach is that, though it still weakly depend on the invocation depth, by virtue of being less computationally intensive, a large number of invocation depth can be covered. Thereby increasing the likelihood of identifying the potential vulnerabilities, which might be embedded deep inside. There are primarily three types of vulnerabilities: (i) Type-1: Suicidal, (ii) Type-2: Prodigal, (iii) Type-3: Greedy [23, 25]. The brief description about these vulnerabilities are given below.

*Suicidal contracts* : A smart contract often has a security fallback option using which the contract can be terminated by a trusted address. This option is enabled to tackle emergency situations such as loss of ether due to attacks or other malfunctioning of the contract. If this feature is not implemented properly, it poses a vulnerability where a contract can be killed by any address. Such contracts are called Suicidal.

*Prodigal contracts* : Smart contracts that can send funds to arbitrary addresses are termed as Prodigal contracts. When under attack, the contracts can send funds to owners or to arbitrary addresses. Attackers can exploit this vulnerability and send Ether to their own accounts.

*Greedy contracts* : A smart contract that cannot release Ether to any address are termed as Greedy contracts. These are further sub divided into two categories: a) contracts that accept Ether but lack the instructions to send funds and b) contracts that accept Ether and contain instructions to send funds, but unable to perform the task.

## 2. METHODS

### 2.1 Data pre-processing

The SC data that is analyzed in this article is obtained from the Tann et al. work [25]. They sourced the original data from Google BigQuery and then removed the false-positives present in it [23], because it is likely to influence the performance. They have used MAIAN to obtain the labels based on vulnerability categories, as well as to extract the opcodes from the bytecodes [16]. This dataset, after preprocessing has 892913 distinct addresses, labelled in five different types, based on the vulnerability categories as (i) Type-1: Suicidal, (ii) Type-2: Prodigal, (iii) Type-3: Greedy [23, 25] (iv) Type-4 normal SCs, and (v) Type-5: Prodigal and Greedy both (The brief description is given in *Section 1.3*). The number of SCs falls under these categories are 5801, 1461, 1207, 884273, and 171 respectively. Out of this, we have selected the first four types (Type-1 to 4). We did not consider the SCs of Type 5 (therefore not considered for further processing) primarily because of the two following reasons: (i) Type-5 is a composite category and therefore we tried avoiding it in this concept paper, and (ii) having too few SCs in this category (only 171 when compared to 884273 for Type-4) will create huge class imbalance, which might be challenging to address even by the strategy that we adopted and this is discussed in the following. Thus we have processed 892742 SCs. As we can see that the negative cases (Type 4: normal SCs) are large in number as compared to the positive cases (Type 1-3: SCs with vulnerabilities). Therefore it posed the problem of class imbalance, which is a serious issue from the vulnerability identification viewpoint. The main reason behind this class imbalance is the presence of repeated opcode combinations. It means that, whenever there is a new invocation of the SC, the new address is appended, whereas the opcode combination remains unchanged. Therefore a single specific opcode combination will refer to multiple instances of addresses. In a general setting, these will be referred to as distinct SCs, whereas we need not treat them as different. We, therefore, can reduce the computational efforts in identifying the vulnerabilities by considering only the distinct combinations of the opcodes that truly represent distinct situations. It also has an added advantage of reducing the adverse class imbalance, thereby further improving the performance. To put the matter in perspective, while we had 892742 distinct addresses, there are only 34822 distinct opcodes combinations. However, to tackle the problem of class imbalance, we have used all the opcodes belonging to the vulnerable SCs and only the distinct opcodes for normal SCs. That is, we did not remove any opcode combinations for Type 1 to 3 (their number remained intact as 5801, 1461 and 1207 respectively). However, for Type 4 (normal category) out of the total 884273 opcode combinations, only the unique combinations which are 32408 were retained and all the duplicate ones deleted. So finally, we have analyzed 40,877 opcode combinations (Type-1 to 4 as 5801, 1461, 1207, and 32408, respectively).

## 2.2 Model: AWD-LSTM

The smart contract data is sequential data, where the input is an opcode sequence. Thus, among various machine learning models, we have used the Average Stochastic Gradient Descent Weight-Dropped Long Short-Term Memory (AWD-LSTM) model [26]. It is one of the most popular state-of-the-art language models and has been widely used in natural language processing [27]. It is ideally suited for sequential data, where some remnant memory of previous constructions lingers to a later stage, which is a foundational characteristic feature of any natural language (and so also is the case for SCs). LSTM model has been applied on different squential data to solve different problems such as natural language processing (NLP) [28,29], speech recognition [30,31], machine translation [32,33], etc. Tann *et al*. [25] used a standard LSTM model [34,35], where the embeddings are initialized randomly at the beginning of training and are updated during training. This study did not make a distinction between different types of vulnerabilities and considered them all in a unified class. In essence, it, therefore, implemented a binary classification: either having a vulnerability or not. The present article, however, attempted a multi-class classification, thereby producing a richer and deeper insight about the SCs. Rather than using a standard LSTM model for classification, we used a combination of two networks. The first network is similar to a Language model used in NLP, where the target task is to predict the next word in a sentence. The second network performs classification using representations learned in the language model.

An LSTM model uses only two distinct weight matrices, also known as embedding matrices which are learned during training of the neural network. The number of rows (say, $m$) of the embedding matrix facing the input layer should be identical to the number of vocabulary, while the number of columns (say, $n$) is, typically, a hyperparameter. The inputs are pre-processed in the form of 'one hot vector' [25] having length m. Therefore, after the multiplication of input vector and the embedding matrix, the output will be of size $1 \times n$, which will be one of the three inputs for LSTM block. The other two inputs (the hidden state and the "memory" of previous block) typically are set to zero for the first LSTM block. An LSTM block consists of four main components : a) Forget gate b) Input gate c) Output gate and d) cell (memory) state. The cell state helps in passing information from one LSTM block to another while learning dependencies in long sequences. The forget gate removes the insignificant information from the cell state by multiplying it with a filter. It takes two inputs: hidden state from the previous time step and the input at the current time step. These inputs are multiplied by the weight matrices followed by a sigmoid activation function which gives output as a vector with values ranging from 0 to 1. If the output for a particular cell state is '0', it means that the forget gate wants the cell state to forget that piece of information completely. The input gate is responsible for adding new information to the cell state. The output gate helps in selecting relevant information from the current cell state. The lack of requirement of fixed shape helps to build neural network models on sequential data with variable length. It is proven to be suitable for tasks such as machine translation where both input and output can have variable length. The embedding matrix has a lot of contextual information about the smart contracts and hence a good embedding matrix is very crucial to achieve better performance. The specific strategy adopted in this study for obtaining a good embedding matrix is discussed in the following.

The AWD-LSTM is primarily used for a particular type of data, where both the inputs and outputs are sequential. The target task is similar to that of language modeling in NLP, where we predict the next word in a sentence given a sequence of input words. Similarly, for the SCs, we train the network to predict the next opcode given a sequence of opcodes. A generic AWD-LSTM primarily consists of two layers: encoder and decoder (Fig. 1). The encoder consists of some LSTM blocks and learns a compact representation of the input, which is later used by the decoder to predict the output. For example, if the input is an opcode sequence that contains elements with indices from 1 to 100, the output is from the same opcode sequence, which contains elements with indices from 2 to 101 (shifted by one index).

First, we train and validate this 'regular architecture' AWD-LSTM with the input and output vectors, both having the same length. The length of the input opcode sequence is of variable length. However, to achieve multi-class classification, the specific adaptation of the algorithm in this study is implemented by replacing the 'decoder' layer of AWD-LSTM by some fully connected layers, also known as custom head. Needless to point that, the output layer of this block is the same as the total number of Types, which is four in this particular study. It is to be noted that, the encoder in first block gets trained during the first phase which will act like a pre-trained encoder for the second block (classifier). It is better than random initialization, and the networks already contains a lot of semantic information about the input data. ***The high-level idea of the present protocol is to combine a pretrained encoder with the 'custom head' to obtain a better classification: this is motivated by ULMFIT as implemented for the NLP application as articulated in [27].*** This modified architecture is trained and validated again. There are other neural network architectures such as Transformer, which is a form of Deep Neural Networks (DNN) and employ the idea of transfer learning to perfrom NLP tasks such as language translation might also be useful.

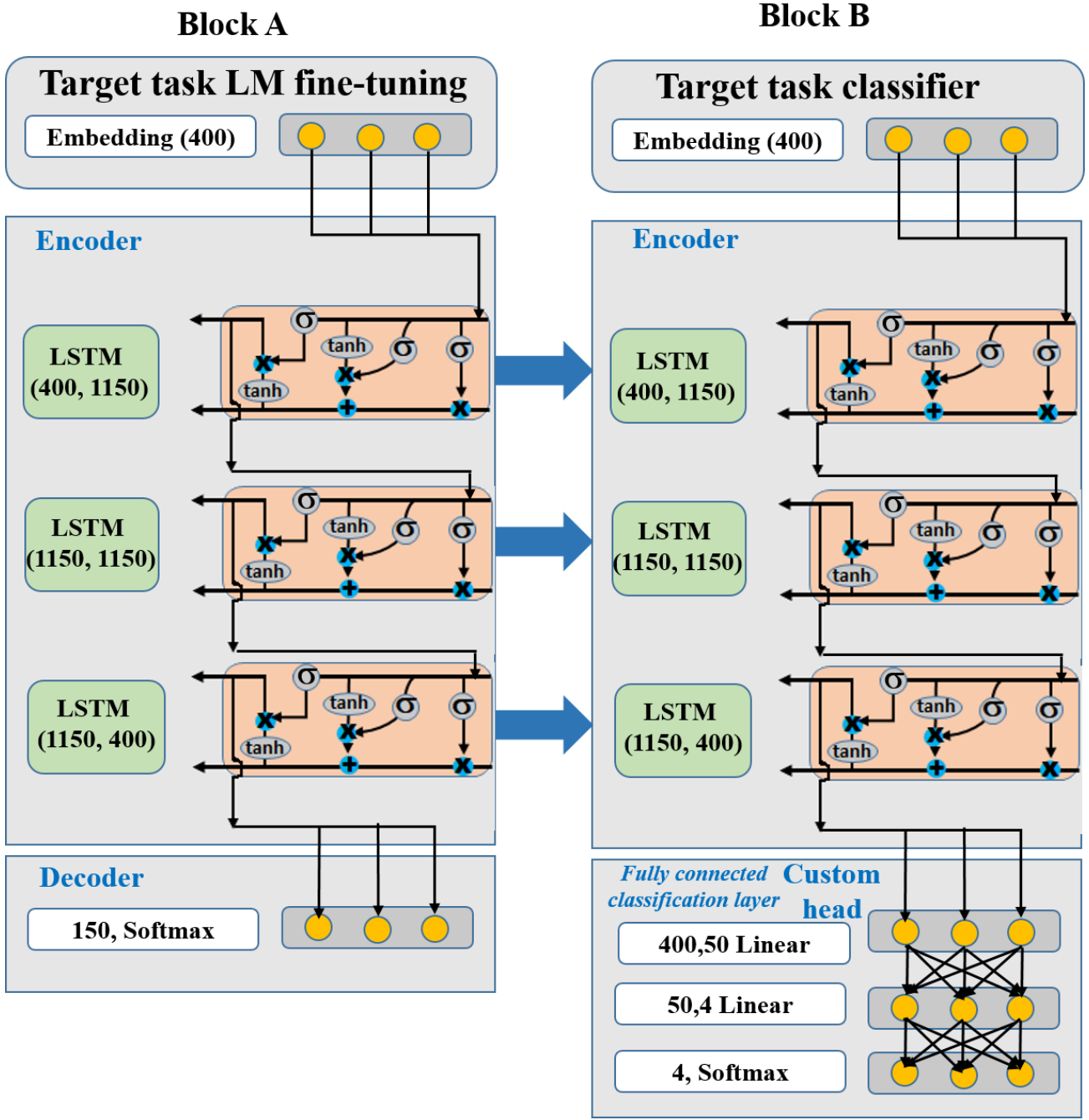

Fig. 1: The architecture of AWD-LSTM used in the present study. Block-A corresponds to the Language model-like architecture and Block-B corresponds to the Classification network. Blue arrows indicate that the weights of the encoder layer of Block-A are copied to the encoder layer of Block-B. The size of input embedding layer is 400. The comprehensive details of this architecture can be found in [36]

### 2.3 Training details

Out of the total 40,877 opcode combinations available after the pre-processing, as discussed in section 2.1, we have divided each class types spanning from 1 to 4 (number of SCs belonging to them are 5801, 1461, 1207, and 32408, respectively) in training, validation and test sets uniformly at the ratio of 70:15:15. So the test cases for each types consists of 870, 220, 181, and 4860 respectively totalling 6131 opcodes. Two neural network blocks (A and B as shown in Fig. 1) are used in the present study. The first network is AWD-LSTM, which is trained to predict the next element in an opcode sequence given a few elements in that opcode sequence. To avoid overfitting, AWD-LSTM(Block-A) is regularized using various types of dropouts [26] such as (a) Embedding dropout where dropout is applied to remove words from the embedding layer for a single epoch which removes the occurrence of a specific word within that epoch, (b) Variational dropout, (c) L2 regularization on the weights which restricts the weights from getting large in magnitude [details can be found in 26]. The second AWD-LSTM (Block-B) network is trained to perform classification.

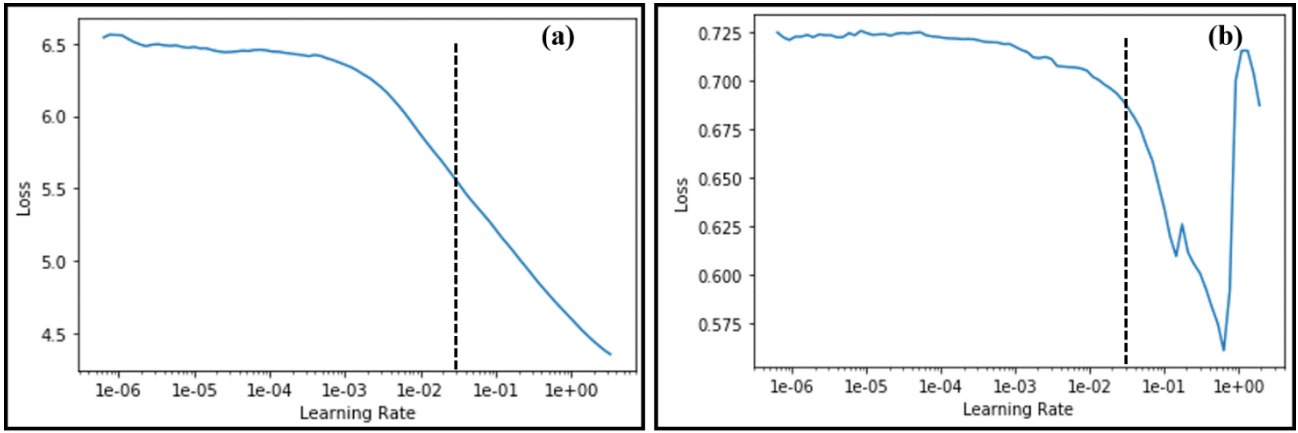

Fig. 2 (a) Learning rate scheduler for Language model (b) Learning rate scheduler for Classifier

We used the learning rate scheduler [37] to find the optimal learning rate for the given data set. In this process, the learning rate is gradually increased after each mini-batch, and the loss is recorded at each increment and loss vs. learning rate is plotted in Fig. 2. For learning rates that are too low, the loss decreases at a slower rate, but as we reach the optimal learning rate region, the loss will drop quickly (steepest slope). Increasing the learning rate beyond this value may result in overshooting the global minima of the loss function. To achieve good results, we train the model with a learning rate that corresponds to the point where the loss reduces most steeply for the first few epochs. Figure 2 indicates that 0.03 is the optimal learning rate for first neural network as well as for the classifier network because after this point the slope of the loss curve is very high. Instead of training the whole network with a single learning rate, discriminative learning rates are used to train the classifier where learning rates lie in the range of (0.0044 to 0.04). Different layers of the network are trained at different learning rates since they capture different types of information. Initial layers of the network are trained at lower learning rates as compared to later layers of the network. During classification, the encoder part of the network is frozen (parameters are not updated) initially for few epochs, and only the custom head is fine-tuned. After the elapse of those few epochs, the encoder part is gradually unfrozen layer by layer, and the network is trained progressively. Both the networks are trained using a one-cycle training policy [37], where the learning rate is high for initial epochs and reduced significantly for the last epoch. These techniques help us in achieving better performance and stable training, where the parameters are updated at the correct pace. This strategy helps us in taking advantage of the knowledge gained from the language model in the form of a pretrained encoder (as articulated in the high-level idea mentioned in section 2.2) and helps in obtaining progressively smooth training. The code is written in Python using fastai [38] which is an open source platform to develop deep learning models. The code is executed in Google Collaboratory which provides K-80 GPU with 12GB RAM for free to run machine learning algorithms. The Source code and the weights of the neural networks used in this study is available here [39].

## 3. RESULTS

We have studied 40,877 opcodes combinations (type-1 to 4 as 5801, 1461, 1207 and 32408 respectively) and analyzed the performance of AWD-LSTM method on the given dataset by calculating the following metrics:

**Accuracy:** The accuracy of the model is calculated based on the number of the correctly identified vulnerability

$$\text{Accuracy} = \frac{\text{True Positives} + \text{True Negatives}}{\text{Total tested}} \qquad (1)$$

The accuracy is not always the best metric for defining the usefulness of the model. For example, when we have very few positive cases in a huge dataset (class imbalanced dataset), even if the model fails to detect any true positive case, the accuracy will remain high (due to the high value of true negatives in Eq. 1). Thus we do not rely only on accuracy but also calculate the recall and precision scores.

**Recall Score:** Recall score tells the ability of a model to find all the relevant or positive cases in the given dataset. It is calculated for each class separately.

$$\text{Recall} = \frac{\text{True Positives}}{\text{True Positives} + \text{False Negatives}} \qquad (2)$$

There is a problem with the recall score, if the model assigns all (or, most of it) the data points as positive cases, then the recall score will be one (or, close to one; Eq. 2). Such a situation will lead to the impression that we have a near-perfect classifier, though in reality, its performance is very poor. To reduce the wrong classification of negative cases, we need a still better metric that maintains the trade-off between true positives and false positives.

**Precision Score:** Precision is the ratio of correctly predicted positive observations to the total predicted positive observations.

$$\text{Precision} = \frac{\text{True Positives}}{\text{True Positives} + \text{False Positives}} \qquad (3)$$

Thus while recall score indicates the ability of a model to find the relevant cases, the precision score tells the proportion of data points, which are categorised as positive cases are actually positive. So from Eqs. (2) and (3), we can see that in a dataset of very few true positives, the recall and precision are inversely proportional. If the recall is high, then precision will be low and vice versa. Thus we need to find the optimal blend of these two metrics.

$F_{beta}$ **Score:** $F_{beta}$ score is the harmonic mean of the precision and recall and can be expressed by Eq. (4).

$$\frac{1}{F_{beta}} = \left( \frac{1}{\text{Precision}} + \frac{1}{\text{Recall}} \right) \Big/ 2 \qquad (4)$$

**Confusion matrix:** Confusion matrix or error matrix also describes the performance of a classification model. It is a square matrix of size $n \times n$, where n is the number of classes present in the dataset. The diagonal elements of the confusion matrix tell the number of correctly classified data points of each class, whereas the other elements are the wrongly classified data points.

**Weighted average of different metrics**

We also calculated the weighted average ($W_m$) of different metrics discussed above using Eq. (5)

$$W_m = \sum_{i=1}^{r_c} \frac{n_i}{N} m_i \qquad (5)$$

Where, $W_m$ is the weighted metric (recall, precision, and $F_{beta}$ score), $r_c$ is the number of categories (four in the present study), $n_i$ is the number of SC of the $i^{th}$ Type, $N$ is the total number of SC and $m_i$ is the respective class metric value (un-weighted).

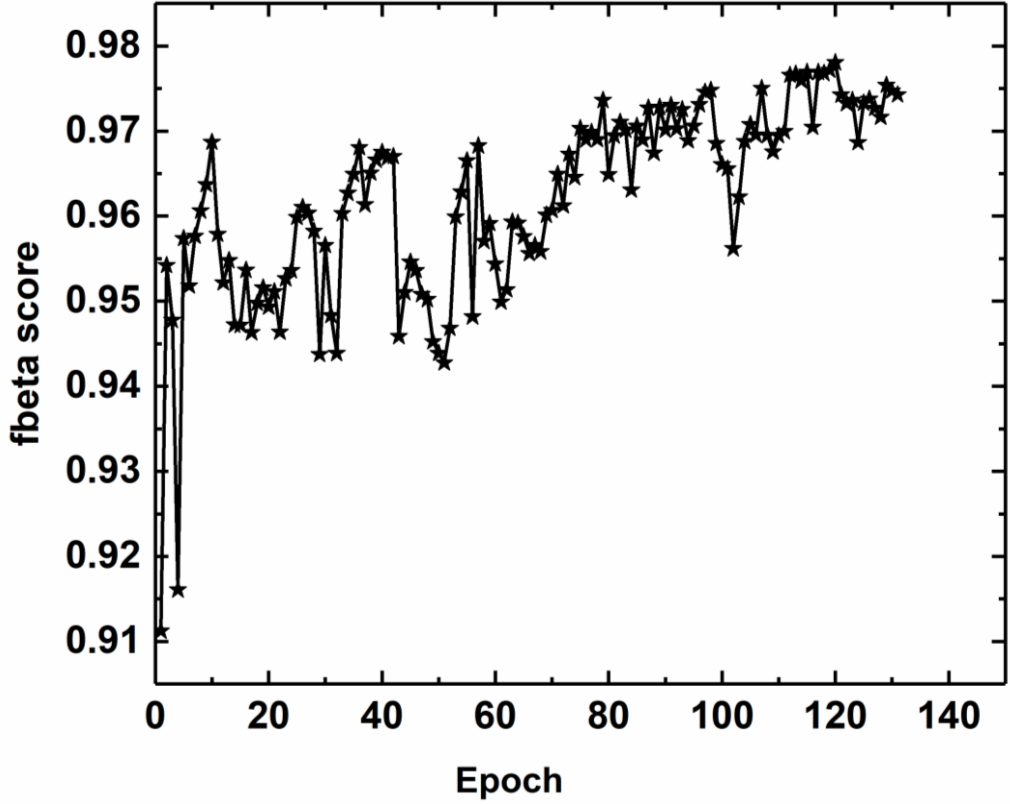

Fig. 3: Variation of $F_{beta}$ score with epochs.

In general, the $F_{beta}$ score increases during training of the model. The variation of $F_{beta}$ score during training is shown in Fig. 3. It shows that the model quickly achieve very high $F_{beta}$ score (in the second epoch itself, while even in the first epoch it is as high as ~91.1%). The model is trained for 132 epochs. The $F_{beta}$ score curve has fluctuations but overall it increases for higher epochs.

Table 1: AWD-LSTM detection performance measures.

| Classification Performance Measure | AWD-LSTM (%) Class-wise | AWD-LSTM (%) Weighted Average |
|---|---|---|
| Test Accuracy | 91.0 (Overall) | --- |
| Recall Score | 74.0 (Type-1) | 91.0 |
|  | 19.0 (Type-2) |  |
|  | 75.0 (Type-3) |  |
|  | 98.0 (Type-4) |  |
| Precision Score | 82.0 (Type-1) | 90.0 |
|  | 66.0 (Type-2) |  |
|  | 94.0 (Type-3) |  |
|  | 93.0 (Type-4) |  |
| $F_{beta}$ Score | 78.0 (Type-1) | 90.0 |
|  | 30.0 (Type-2) |  |
|  | 83.0 (Type-3) |  |
|  | 95.0 (Type-4) |  |

Table 1 shows the values of different metrics used for evaluating the performance of AWD-LSTM. We achieved a higher weighted average $F_{beta}$ score than obtained by [25].

The diagonal elements of the confusion matrix, C (depicted in Fig. 4), are the correctly classified SCs. The elements $C_{1,1}$, $C_{2,2}$, $C_{3,3}$, $C_{4,4}$, are the number of SCs that are correctly classified as Suicidal, Prodigal, Greedy, and normal, respectively. The other elements are miss-classified SCs. The confusion matrix depicts that our model in correctly classifying a minimum of 74% for any Type except for the Type 2. Therefore, it is clear that the prodigal type of vulnerability (class-2), is evading appropriate detection. It is presently unclear, whether this is because of some technical nature of the Type-2 vulnerabilities or becase of some kind of weakness in the present scheme. This, therefore deserves a thorough further investigation.

|  | | Actual | | | |
|---|---|---|---|---|---|
|  | | Type-1 | Type-2 | Type-3 | Type-4 |
| Predicted | Type-1 | 648 | 4 | 3 | 215 |
|  | Type-2 | 59 | 42 | 0 | 119 |
|  | Type-3 | 1 | 0 | 135 | 45 |
|  | Type-4 | 78 | 18 | 5 | 4759 |

Fig. 4: (a) Confusion matrix for all the four vulnerable categories for the total 6131 opcodes belonging to the validation and test sets (for details, see section 2.3).

Receiver Operating Characteristics (ROC) curve, shown in Fig. 5, is the plot between true positive rate and false positive rate of predictions from a neural network for various classes studied in this article. In the limit of AUC (Area under curve) approaching to one, the model will be close to perfection and will indicate near ideal classification. A value closer to 0.5 will indicate detereorating performance. The AUC metric (mentioned in the legend in Fig. 5) indicates the neural networks' capability to distinguish between various classes. The best performance is observed for Type-3 (AUC >99%), while the minimum, which is observed for Type-2 (AUC >98%) is still very good. Therefore, we can reasonably conclude that the method proposed in this article for multi-class classification of SCs is fairly accurate, acceptable and can provide a foundation for further development.

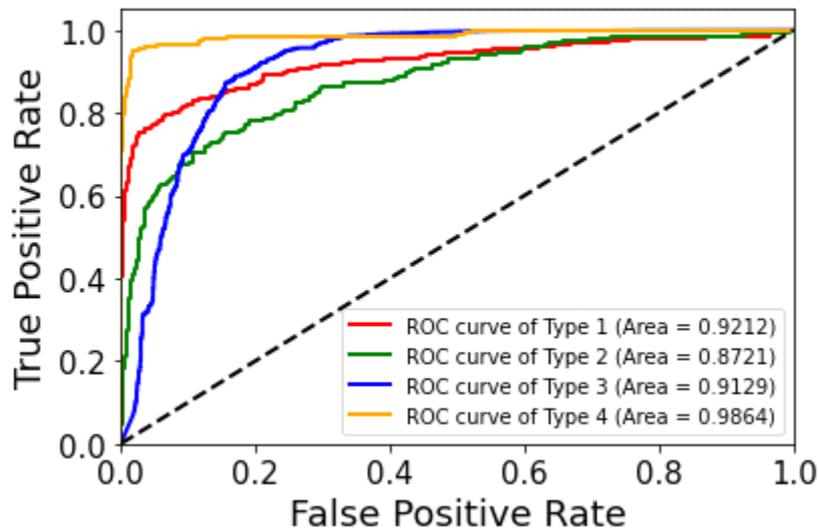

Fig. 5: Receiver operating characteristics (ROC) curve

## Conclusion

The adaptation of Pre-trained neutral networks is increasing in diverse areas of deep-learning applications as they are proven to be useful in achieving better performance in various tasks. It was initially used for image classification and is extended to various tasks in NLP. A neural network is trained initially on a different target task for which a large amount of data is available. Later, changes are made to the neural network architecture by replacing some blocks in the initial network to perform the required target task. The high-level idea of the present protocol is to combine a pretrained encoder with the 'custom head' to obtain a better classification: this is motivated by ULMFIT. We adapted this protocol for multi class classification for the SCs. In particular, we used two neural networks where the first network learns a significant amount of semantic information about the input data which helps the second network to achieve better and quicker performance. We demonstrated that the method proposed in this article for multi-class classification of SCs is fairly accurate and produces acceptable results with an accuracy of 91.0% and an $F_{beta}$ score of 90.0%. Another important metric, AUC is also very high (minimum value is in excess of 0.87 for any class), indicating robust performance of the present algorithm for the detection of vulnerabilities in the SCs. We also outlined the scope and direction for future research for improved performance.

## Acknowledgement

We thank Sourav Sen Gupta, Wesley Joon-Wie Tann, Xing Jie Han and, Yew-Soon Ong for sharing the preprocessed SC data with us, which was used for this study.

## References

1. Szabo, N. (1996). Smart contracts: building blocks for digital markets. *EXTROPY: The Journal of Transhumanist Thought,(16)*, *18*, 2.
2. Antonopoulos, A. M., & Wood, G. (2018). Mastering ethereum: building smart contracts and dapps. O'reilly Media.
3. The Ultimate Guide to Rootstock Blockchain.
   https://blockgeeks.com/guides/rootstock-blockchain/
4. Buterin, V.: Ethereum: a next generation smart contract and decentralized application platform. https://github.com/ethereum/wiki/wiki/White-Paper (2013).

5. Hacking Distributed. 2016. Analysis of the DAO exploit. http://hackingdistributed.com/2016/06/18/analysis-of-the-dao-exploit/. [Online; accessed 20-November-2019].
6. Parity Technologies. 2017. Security Alert. https://paritytech.io/security-alert-2/.[Online; accessed 20-Novemebr-2019].
7. Erickson, B. J., Korfiatis, P., Akkus, Z., & Kline, T. L. (2017). Machine learning for medical imaging. Radiographics, 37(2), 505-515.
8. Huang, X., Shan, J., & Vaidya, V. (2017, April). Lung nodule detection in CT using 3D convolutional neural networks. In 2017 IEEE 14th International Symposium on Biomedical Imaging (ISBI 2017) (pp. 379-383). IEEE.
9. Ishii, T., Nakamura, R., Nakada, H., Mochizuki, Y., & Ishikawa, H. (2015, May). Surface object recognition with CNN and SVM in Landsat 8 images. In 2015 14th IAPR International Conference on Machine Vision Applications (MVA) (pp. 341-344). IEEE.
10. Liu, Y., Zhao, T., Ju, W., & Shi, S. (2017). Materials discovery and design using machine learning. Journal of Materiomics, 3(3), 159-177.
11. Liu, R., Kumar, A., Chen, Z., Agrawal, A., Sundararaghavan, V., & Choudhary, A. (2015). A predictive machine learning approach for microstructure optimization and materials design. Scientific reports, 5, 11551.
12. Li, X., Xie, H., Wang, R., Cai, Y., Cao, J., Wang, F., ... & Deng, X. (2016). Empirical analysis: stock market prediction via extreme learning machine. Neural Computing and Applications, 27(1), 67-78.
13. Du, M., Li, F., Zheng, G., & Srikumar, V. (2017, October). Deeplog: Anomaly detection and diagnosis from system logs through deep learning. In Proceedings of the 2017 ACM SIGSAC Conference on Computer and Communications Security (pp. 1285-1298). ACM.
14. Shen, Y., Mariconti, E., Vervier, P. A., & Stringhini, G. (2018, October). Tiresias: Predicting security events through deep learning. In Proceedings of the 2018 ACM SIGSAC Conference on Computer and Communications Security (pp. 592-605). ACM.
15. Shin, E. C. R., Song, D., & Moazzezi, R. (2015). Recognizing functions in binaries with neural networks. In 24th {USENIX} Security Symposium ({USENIX} Security 15) (pp. 611-626).
16. Wood, G. (2014). Ethereum: A secure decentralised generalised transaction ledger. Ethereum project yellow paper, 151(2014), 1-32.
17. Six Popular Blockchain Programming Languages Used for Building Smart Contracts — And FLETA will support them all. https://medium.com/fleta-first-chain/6-popular-blockchain-programming-languages-used-for-building-smart-contracts-and-fleta-will-7b310f1a9e2
18. Bistarelli, S., Mazzante, G., Micheletti, M., Mostarda, L., & Tiezzi, F. (2019, March). Analysis of Ethereum Smart Contracts and Opcodes. In International Conference on Advanced Information Networking and Applications (pp. 546-558). Springer, Cham.
19. Chen, W., Zheng, Z., Cui, J., Ngai, E., Zheng, P., & Zhou, Y. (2018, April). Detecting ponzi schemes on ethereum: Towards healthier blockchain technology. In Proceedings of the 2018 World Wide Web Conference (pp. 1409-1418). International World Wide Web Conferences Steering Committee.
20. Luu, L., Chu, D. H., Olickel, H., Saxena, P., & Hobor, A. (2016, October). Making smart contracts smarter. In Proceedings of the 2016 ACM SIGSAC conference on computer and communications security (pp. 254-269). ACM.
21. "Oyente: An Analysis Tool for Smart Contracts," 2018. [Online]. Available: https://github.com/melonproject/oyente
22. MAIAN: automatic tool for finding trace vulnerabilities in ethereum smart contracts. https://github.com/MAIAN-tool/MAIAN. Accessed 22 January 2020.
23. Nikolić, I., Kolluri, A., Sergey, I., Saxena, P., & Hobor, A. (2018, December). Finding the greedy, prodigal, and suicidal contracts at scale. In Proceedings of the 34th Annual Computer Security Applications Conference (pp. 653-663). ACM.

24. Liu, C., Liu, H., Cao, Z., Chen, Z., Chen, B., & Roscoe, B. (2018, May). Reguard: finding reentrancy bugs in smart contracts. In Proceedings of the 40th International Conference on Software Engineering: Companion Proceeedings (pp. 65-68). ACM.
25. Tann, A., Han, X. J., Gupta, S. S., & Ong, Y. S. (2018). Towards safer smart contracts: A sequence learning approach to detecting vulnerabilities. arXiv preprint arXiv:1811.06632.
26. Merity, S., Keskar, N. S., & Socher, R. (2017). Regularizing and optimizing LSTM language models. arXiv preprint arXiv:1708.02182.
27. Howard, J., & Ruder, S. (2018). Universal language model fine-tuning for text classification. arXiv preprint arXiv:1801.06146.
28. Graves, A. (2013). Generating sequences with recurrent neural networks. arXiv preprint arXiv:1308.0850.
29. Luong, M. T., Pham, H., & Manning, C. D. (2015). Effective approaches to attention-based neural machine translation. arXiv preprint arXiv:1508.04025.
30. Dahl, G. E., Yu, D., Deng, L., & Acero, A. (2011). Context-dependent pre-trained deep neural networks for large-vocabulary speech recognition. IEEE Transactions on audio, speech, and language processing, 20(1), 30-42.
31. Hinton, G., Deng, L., Yu, D., Dahl, G. E., Mohamed, A. R., Jaitly, N., ... & Kingsbury, B. (2012). Deep neural networks for acoustic modeling in speech recognition: The shared views of four research groups. IEEE Signal processing magazine, 29(6), 82-97.
32. Cho, K., Van Merriënboer, B., Gulcehre, C., Bahdanau, D., Bougares, F., Schwenk, H., & Bengio, Y. (2014). Learning phrase representations using RNN encoder-decoder for statistical machine translation. arXiv preprint arXiv:1406.1078.
33. Wu, Y., Schuster, M., Chen, Z., Le, Q. V., Norouzi, M., Macherey, W., ... & Klingner, J. (2016). Google's neural machine translation system: Bridging the gap between human and machine translation. arXiv preprint arXiv:1609.08144.
34. Hochreiter, S., & Schmidhuber, J. (1997). Long short-term memory. Neural computation, 9(8), 1735-1780.
35. Gers, F. A., Schmidhuber, J., & Cummins, F. (1999). Learning to forget: Continual prediction with LSTM.
36. Sandra F. et al. Universal Language Model Fine-Tuning (ULMFiT) State-of-the-Art in Text Analysis https://humboldt-wi.github.io/blog/research/information_systems_1819/group4_ulmfit/ [Online; accessed 22-November-2019].
37. Smith, L. N. (2017, March). Cyclical learning rates for training neural networks. In 2017 IEEE Winter Conference on Applications of Computer Vision (WACV) (pp. 464-472). IEEE.
38. NLP model creation and training, Weblink: https://docs.fast.ai/text.learner.html [Online; accessed 22-January-2020].
39. https://github.com/AjayKumarGogineni777/Smart-Contract-analysis-using-AWD-LSTM